\documentclass[prc,twocolumn,showpacs,showkeywords,showpreprintnumbers,floatfix]{revtex4}
\usepackage{amsmath,amssymb,graphicx,bm}
\usepackage{color}
\usepackage{ulem}

\newcommand{\vek}[1]{\bm{\mathrm{#1}}}
\DeclareMathOperator{\tr}{Tr}
\DeclareMathOperator{\im}{Im}
\DeclareMathOperator{\re}{Re}
\newcommand{\vlowk}{V_{{\rm low}\,k}}
\newcommand{\fm}{\, \text{fm}}
\newcommand{\fmi}{\, \text{fm}^{-1}}
\newcommand{\fmitwo}{\, \text{fm}^{-2}}
\newcommand{\fmithree}{\, \text{fm}^{-3}}
\newcommand{\MeV}{\, \text{MeV}}
\newcommand{\vekp}{\vek{p}}
\newcommand{\vekq}{\vek{q}}
\newcommand{\vekk}{\vek{k}}
\newcommand{\vekK}{\vek{K}}
\newcommand{\Gtemp}{\mathcal{G}}
\newcommand{\avGtemp}{\overline{\Gtemp}}
\newcommand{\avG}{\overline{G}}
\newcommand{\avQ}{\overline{Q}}
\newcommand{\rhocorr}{\rho_{\rm corr}}
\newcommand{\rhofree}{\rho_{\rm free}}

\begin{document}

\title{BEC-BCS Crossover in Neutron Matter with
  Renormalization Group based Effective Interactions}
\author{S. Ramanan} \email{suna@physics.iitm.ac.in}
\affiliation{Department of Physics, Indian Institute of Technology Madras,
  Chennai - 600036, India} \author{M. Urban}
\email{urban@ipno.in2p3.fr} \affiliation{Institut de Physique
  Nucl\'eaire, CNRS-IN2P3 and Universit\'e Paris-Sud, 91406 Orsay
  Cedex, France} 
  %\date{July 26, 2013}

\begin{abstract}
We study pure neutron matter in the BEC-BCS crossover regime using
renormalization group based low-momentum interactions within the
Nozi\`eres-Schmitt-Rink framework. This is an attempt to go beyond the
mean field description for low-density matter. We work in the basis of
so-called Weinberg eigenvectors where the operator $G_0V$ is diagonal,
which proves to be an excellent choice that allows one to use
non-local interactions in a very convenient way. We study the
importance of correlations as a function of density. We notice that
there is a significant reduction of the BCS critical temperature at
low-densities as the neutron matter approaches the unitary limit.
\end{abstract}



\maketitle

\section{Introduction}
\label{sect:intro}

The study of pairing correlations between nucleons is important for
both infinite matter such as neutron stars as well as for finite
nuclei, especially close to the drip lines~\cite{Margueron:2007uk,
  Baldo:2004dx, Monrozeau:2007xu, Yakovlev:2004iq, Hagino:2006ib,
  Hagino:2005we, Sagawa:2007we, Sagawa:2012je}. In neutron matter at
low density, strong correlations build up between the interacting pair
of neutrons due to weak Pauli-blocking~\cite{Baldo:2004dx}. Evidence
for such correlations have been observed in nuclei close to the drip
lines, for example in $^{11}$Li where the two neutrons outside the
core become strongly correlated~\cite{Hagino:2006ib, Hagino:2005we,
  Sagawa:2007we, Sagawa:2012je}. In addition to playing a crucial role
in halo nuclei, strong correlations are important to explain the
glitches as well as the cooling rates of neutron
stars~\cite{Monrozeau:2007xu, Yakovlev:2004iq}. Typically a neutron
star, which is born at the end of a core collapse supernova, consists
of asymmetric nuclear matter, i.e. neutrons and protons in $\beta$
equilibrium with the electrons. In the inner crust, protons and
neutrons form clusters that are surrounded by a superfluid neutron
gas.

In this paper we study pure neutron matter and analyze the pairing
correlations as a function of density. In symmetric matter at low
density, one expects a Bose-Einstein condensate (BEC) of
deuterons~\cite{rostock1,rostock2,Jin:2010nj}, but in pure neutron
matter, low density does not lead to a BEC state, because there is no
bound di-neutron state. Nevertheless it results in matter being
strongly correlated. In fact, at low densities neutron matter tends
almost to a unitary gas as introduced by Bertsch \cite{Baker1999}
because of the unusually large neutron-neutron scattering length of
$\approx -18.7$ fm \cite{Gonzalez1999}. At higher density, the pair
correlations become less important and the system can be described
within Bardeen-Cooper-Schrieffer (BCS) mean-field theory. Pairing in
neutron matter has been widely studied within the BCS theory (for an
overview, see, e.g., Chapter 8.2 of \cite{ChamelHaensel2008}), but in
the strongly correlated regime the transition temperature $T_c$ is
strongly overestimated within this approach. In order to go beyond the
mean field approximation, we need to include pair correlations above
$T_c$. The Nozi\`eres-Schmitt-Rink (NSR) approach involves the
inclusion of two-particle correlations into the density above $T_c$
within the ladder approximation~\cite{NSR}. At higher density, a very
similar approach by the Rostock group \cite{rostock1,rostock2},
building the ladder diagrams out of quasiparticles instead of free
particles, seems to be more adequate.

In the present work, we will use renormalization group based effective
interactions ($\vlowk$) as the two-body input (for a recent review, we
refer the reader to ~\cite{Bogner_rev2009}). Such interactions were
already employed to describe pairing in finite nuclei and infinite
matter \cite{Hebeler2007,Lesinski2009,Hebeler2009}. They allow for a
model independent approach to the nuclear many-body problem and one
can use the residual dependence on the renormalization scale as a tool
to get an estimate of the missing many-body physics. These
interactions are non-local. For a general potential $V$, the magnitude
of the eigenvalues of the operator $G_0 V$, where $G_0$ is the
two-body Green's function, quantifies the convergence of a Born series
expansion of the $T$ matrix. If there are sources of non-perturbative
physics, these show up in the so-called Weinberg eigenvalues of this
operator~\cite{Weinberg,bogner2006,Ramanan:2007bb}. For example, in
free space a bound state corresponds to a pole in the $T$ matrix. The
Weinberg eigenvalue equation for the operator $G_0 V$ at the binding
energy is just the Schr\"odinger equation for a bound state and hence
the eigenvalue equals $1$. In the many-body system, the formation of
Cooper pairs is very similar to the formation of bound states. This
idea has been previously used to obtain the BCS pairing gap by looking
at the eigenvalues crossing $1$ close to the Fermi
surface~\cite{Ramanan:2007bb}. In this paper we generalize this idea
to finite temperature, which is the Thouless criterion for the onset
of pairing~\cite{NSR,thouless,RingSchuck}. In addition to obtaining
the transition temperature, the basis where the operator $G_0 V$ is
diagonal offers additional simplification for the calculation of the
correlated density.

This paper has been organized as follows. In section~\ref{sect:T_c} we
discuss the finite-temperature generalization for the Weinberg
eigenvalue equation and use the eigenvalues to calculate the BCS
critical temperature. In section~\ref{sect:NSR} we obtain a formula
for the correlated density within the NSR approach. We present our
results in section~\ref{sect:results} and discuss the critical
temperature as a function of density and we discuss the implications
of our results for the understanding of pairing correlations in
low-density neutron matter. In section~\ref{sect:summary}, we
  summarize and give an outlook to improvements of the theory that
  should be addressed in future studies.

To simplify the notation, we use units with $\hbar = c = m = k_B = 1$,
where $\hbar$ is the reduced Planck constant, $c$ is the speed of
light, $m$ is the neutron mass, and $k_B$ is the Boltzmann
constant. To convert energies from fm$^{-2}$ into MeV one therefore
has to multiply them by $\hbar^2/m = 41.44 \MeV \fm^2$, and to convert
momenta from fm$^{-1}$ into MeV one has to use $\hbar c = 197.3
\MeV\fm$.
%%%%%%%%%%%%%%%%%%%%%%%%%%%%%%%%%%%%%%%%%%%%%%%%%%%%%%%%%%%%%%%%%%%%%%%%
\section{Finite-temperature eigenvalues and critical temperature}
\label{sect:T_c}
%%%%%%%%%%%%%%%%%%%%%%%%%%%%%%%%%%%%%%%%%%%%%%%%%%%%%%%%%%%%%%%%%%%%%%%%
In this study we are interested in the transition from the normal to
the paired state. The critical temperature that determines the on-set
of pairing correlations can be obtained by the Thouless criterion,
which states that if the $T$ matrix does not exhibit a pole, the
temperature is above the superfluid transition
temperature~\cite{thouless,RingSchuck}. Therefore we look for the
poles of the in-medium finite-temperature $T$ matrix. We do this using
the eigenvalues of the operator $G_0 V$, where $G_0$ is the
non-interacting two-body Green's function~\cite{Weinberg, bogner2006,
  Ramanan:2007bb}.

Let us briefly show how this works in free space. We begin by
considering for a given energy $E$ a basis $\{|\Gamma(E)
\rangle\}$ such that the operator $G_0(E) V$ is diagonal, i.e.,
\begin{equation} 
G_0(E) V |\Gamma(E) \rangle = \eta(E) |\Gamma(E)\rangle, 
\label{eq:wein_op}
\end{equation} 
where $\eta(E)$ is the
eigenvalue~\cite{Weinberg,fewbody,bogner2006}. The form of the Green's
function determines the properties of the eigenvalue. In free space at
zero temperature, for example, we have $G_0(E) = (E-H_0)^{-1} =
(E-K^2/4-q^2)^{-1}$, where $H_0$ is the non-interacting hamiltonian
and $K$ and $q$ are the total and relative momenta of the two
particles, respectively. Then the eigenvalues are complex for positive
energies (to be precise, for $E > K^2/4$) and real for negative
energies. The eigenvalue equation (\ref{eq:wein_op}) can be
inverted to yield the Schr\"odinger equation:
\begin{equation}
\left(H_0 + \frac{V}{\eta(E)}\right) |\Gamma \rangle = E |\Gamma \rangle.
	\label{eq:schr}
\end{equation}
Therefore the eigenvalues have another interpretation: it is the
factor by which the potential should be scaled in order to support a
bound state at the given energy $E$. If the potential does allow a
bound state at an energy $E_b$, then the corresponding eigenvalue
$\eta(E_b)$ equals $1$.

In~\cite{bogner2006,Bogner:2005sn}, the eigenvalues were used as a
diagnostic to track the ``perturbativeness'' of the particle-particle
ladders, both in free space and in-medium, when the renormalization
group based interactions $\vlowk$ were used as inputs,
and~\cite{Ramanan:2007bb} extended the eigenvalue framework to study
pairing at zero temperature. Since we need the transition temperature,
the first step is to generalize the eigenvalue equation
(\ref{eq:wein_op}) to finite temperature $T = 1/\beta$.

We will use the Matsubara (imaginary time) formalism
\cite{Fetter_walecka}, in which the non-interacting single-particle
Green's function is given by
\begin{equation}
\Gtemp_0^{(1)} (\vekk, \omega_n) = \frac{1}{i \omega_n - \epsilon_{\vekk} + \mu},
\label{eq:sing_gf}
\end{equation}
where $\omega_n = (2n+1)\pi/\beta$ is a fermionic Matsubara frequency
and $\epsilon_{\vekk}$ is the single-particle energy. The
corresponding two-body Green's function is obtained from
\begin{multline}
\Gtemp_0^{(2)}(\vekK, \vekq, \omega_N)
   = -\frac{1}{\beta} \sum_{\omega_n} 
  \Gtemp_0^{(1)}\Big(\frac{\vekK}{2} + \vekq, \omega_n\Big)\\ 
  \times \Gtemp_0^{(1)}\Big(\frac{\vekK}{2} - \vekq, \omega_N - \omega_n\Big)
\label{eq:two_body_gf}
\end{multline}
where $\omega_N = 2\pi N/\beta$ is a bosonic Matsubara frequency.
Using standard techniques \cite{Fetter_walecka} one can perform the
sum over $\omega_n$, and the retarded two-body Green's function at
finite density and temperature for real energy $\omega$ is obtained by
analytic continuation, $i\omega_N\to\omega+i\eta$: 
\begin{equation} 
G_0(\vekK, \vekq, \omega) =
  \frac{1 - f(\xi_{\vekK/2 + \vekq}) - f(\xi_{\vekK/2 - \vekq})}
    {\omega - \xi_{\vekK/2 + \vekq} - \xi_{\vekK/2 - \vekq} + i \eta}.
\label{eq:two_body_gf_retarded}
\end{equation}
where $f(\xi) = 1/(e^{\beta\xi}+1)$ is the Fermi-Dirac distribution
function and $\omega = E-2\mu$, $\xi_{\vekk} = \epsilon_{\vekk}-\mu$
are the energies measured from the chemical potential.

Working in momentum space in a partial wave basis, and considering
only the $^1S_0$ channel, we can write the generalization of the
eigenvalue equation to finite temperature as
follows: 
\begin{multline}
\frac{2}{\pi} \int_0^{q_{max}} dq' q'^2 v(q,q') 
  \avG_0(K, q',\omega)\Gamma(q',K,\omega)\\ 
  = \eta(K,\omega) \Gamma(q,K,\omega)\,,
\label{eq:wein_kspace}
\end{multline}
where $v(q,q')$ is the matrix element of the interaction in the
$^1S_0$ channel and $\avG_0$ is the angle average of $G_0$.  Note that
in equation (\ref{eq:wein_kspace}) we solve for $V G_0$ instead of
$G_0V$ in equation (\ref{eq:wein_op}) analogous to \cite{bogner2006,
  Ramanan:2007bb}. Both operators have the same eigenvalue
spectrum, but the choice $V G_0$ is more convenient in numerical
  calculations since it allows for direct integration over
singularities. In the case of a free particle spectrum,
$\epsilon_{\vekk} = k^2/2$, the angle average can be done analytically
with the result
\begin{equation}
  \avG_0(K,q,\omega) =  
  \frac{\avQ(K, q)}{\omega+2\mu-K^2/4-q^2+i\eta}\,,
\label{eq:avG0}
\end{equation}
where the angle-averaged finite-temperature Pauli-blocking factor
$\avQ(K, q)$ is given by
\begin{equation}
\avQ(K, q) = \frac{2 T}{K q} 
  \log\left(\frac{1 + e^{\beta[(K/2 + q)^2 - 2 \mu]/2}}
    {1 + e^{\beta[(K/2 - q)^2 - 2 \mu]/2}} \right) - 1\,.
\label{eq:ang_av_Q}
\end{equation}

Pairing usually manifests itself as an instability in the two-particle
Green's function or the $T$ matrix~\cite{Ramanan:2007bb,NSR,thouless}. At
zero temperature, the eigenvalues for $K=0$ diverge as $E \rightarrow
2 \mu$ and cross $1$ at energy $E = 2\mu \pm \Delta$, where $\Delta$
is the zero-temperature pairing
gap~\cite{Ramanan:2007bb}. Analogously, at finite temperature, setting
the momentum of the center of mass $K = 0$ and the energy $E = 2\mu$,
one expects at least one eigenvalue to cross $1$ in
Eq.~(\ref{eq:wein_kspace}). The temperature corresponding to such a
crossing is the critical temperature $T_c$ marking the on-set of the
transition from the normal phase to the paired phase.

%%%%%%%%%%%%%%%%%%%%%%%%%%%%%%%%%%%%%%%%%%%%%%%%%%%%%%%%%%%%%%%%%%%%%%%%%
\begin{figure}
\begin{center}
\includegraphics[angle = 0, width = 7.5cm, clip = true]
  {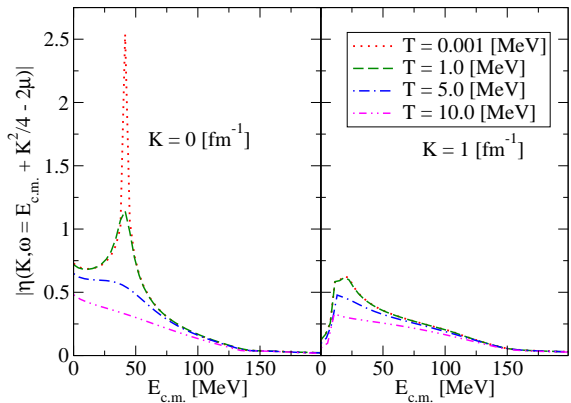}
\end{center}
\caption{(color online) Finite temperature Weinberg eigenvalues:
  Largest eigenvalue as a function of center of mass energy $E_{cm} =
  E - K^2/4$ for different temperatures $T$ and fixed $\mu = 0.5
  \fmitwo$. The left panel corresponds to a center-of-mass momentum $K
  = 0 \fmi$, the right one to $K = 1.0 \fmi$.}
\label{fig:eigen_vs_E}
\end{figure}
%%%%%%%%%%%%%%%%%%%%%%%%%%%%%%%%%%%%%%%%%%%%%%%%%%%%%%%%%%%%%%%%%%%%%%%%%
%%%%%%%%%%%%%%%%%%%%%%%%%%%%%%%%%%%%%%%%%%%%%%%%%%%%%%%%%%%%%%%%%%%%%%%%%
\begin{figure}
\begin{center}
\includegraphics[angle = 0, width = 7.5cm, clip = true]
  {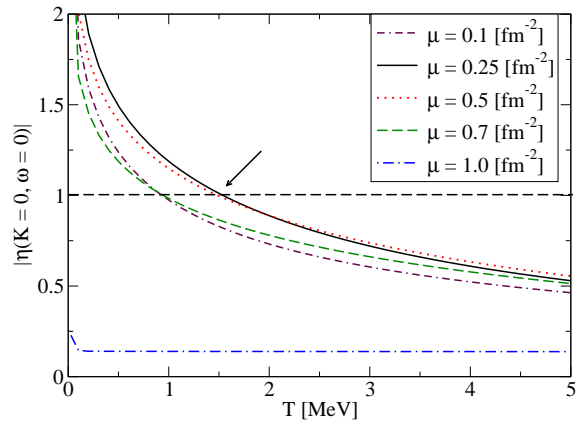}
\end{center}
\caption{(color online) Finite temperature Weinberg eigenvalues:
  Largest eigenvalue for $E=2\mu$ and $K=0$ as a function of $T$ for
  different values of the chemical potential $\mu$. The temperature
  where the eigenvalue crosses $1$ (horizontal dashed line)
  corresponds to the critical temperature $T_c$.}
\label{fig:eigen_vs_T}
\end{figure}
%%%%%%%%%%%%%%%%%%%%%%%%%%%%%%%%%%%%%%%%%%%%%%%%%%%%%%%%%%%%%%%%%%%%%%%%%
This is illustrated in Figs.~\ref{fig:eigen_vs_E} and
\ref{fig:eigen_vs_T}. In our calculations, we use matrix elements
$v(q,q')$ from \cite{Bogner:2006vp} that were obtained from the
AV$_{18}$ interaction with a smooth regulator (F.D. regulator
$\epsilon = 0.5$) at a cutoff of $\Lambda = 2\fmi$ unless otherwise
stated.

Fig.~\ref{fig:eigen_vs_E} shows the magnitude of the largest finite
temperature eigenvalue as a function of center of mass energy for
different values of the center of mass momentum $K$. Note that for low
temperatures, when $K = 0$, the eigenvalue shows a singular behavior
close to the Fermi surface, i.e., at $E=2\mu$, which signals the
pairing instability. Increasing the temperature $T$ or the total
momentum $K$, the Pauli blocking factor gets smeared and the
eigenvalues are no longer singular. The largest Weinberg eigenvalue as
a function of temperature for energy $E = 2 \mu$ and center of mass
momentum $K = 0$ is shown in Fig.~\ref{fig:eigen_vs_T}. The
temperature at which the eigenvalue equals $1$ is the critical
temperature (for example, the transition temperature for $\mu = 0.25
\fmitwo$ is indicated by an arrow in the figure).

\begin{figure}
\includegraphics[angle = 0, clip = true, width = 7.5cm]
  {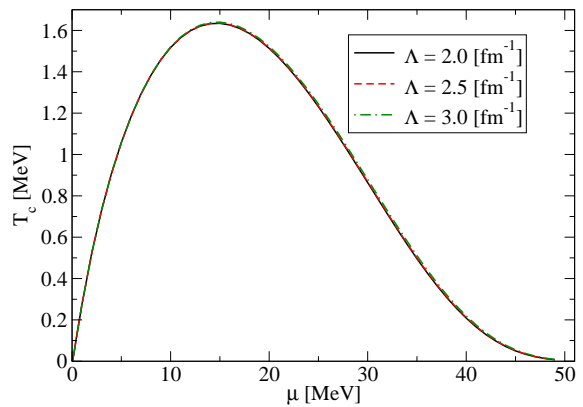}
\caption{(color online) Transition temperature as a function of $\mu$
  obtained from the pole in the $T$ matrix, for different values of
  the cutoff ($\Lambda = 2$, $2.5$ and $3 \fmi$).}
  \label{fig:trans_T_vs_mu}
\end{figure}

Fig.~\ref{fig:trans_T_vs_mu}
%%%%%%%%%%%%%%%%%%%%%%%%%%%%%%%%%%%%%%%%%%%%%%%%%%%%%%%%%%%%%%%%%%%%%%%%%
%%%%%%%%%%%%%%%%%%%%%%%%%%%%%%%%%%%%%%%%%%%%%%%%%%%%%%%%%%%%%%%%%%%%%%%%%
shows the transition temperature $T_c$, determined from the Weinberg
eigenvalues, as a function of $\mu$ for different cutoffs $\Lambda$
between 2 and $3\fmi$. The observed cutoff independence is probably a
peculiarity of the present approach to calculate $T_c$ from the ladder
diagrams with a free single-particle spectrum, the only many-body
effect being Pauli blocking that is not affected by the the truncation
of the model space (as long as the cutoff is above $k_F$).

In the next section, we set up the correlated density in the Weinberg
eigenvector basis and this allows one to generalize the NSR approach
to non-local and non-separable interactions.

%%%%%%%%%%%%%%%%%%%%%%%%%%%%%%%%%%%%%%%%%%%%%%%%%%%%%%%%%%%%%%%%%%%%%%%%%%
\section{Weinberg Eigenvalues and the Nozi\`eres-Schmitt-Rink approach}
\label{sect:NSR}
%%%%%%%%%%%%%%%%%%%%%%%%%%%%%%%%%%%%%%%%%%%%%%%%%%%%%%%%%%%%%%%%%%%%%%%%%%
The NSR approach \cite{NSR} accounts for the fact that there can be
strong pair correlations aleady above the transition temperature
$T_c$. Although there is no bound di-neutron, the presence of the
surrounding medium induces strong correlations and one sometimes
speaks of ``preformed pairs'' that are similar to Cooper pairs, but
not yet condensed into the $K=0$ state. A very similar approach was
developed by the Rostock group~\cite{rostock1,rostock2}, where it was
pointed out that one should think of the entire picture in terms of
free and correlated quasiparticles instead of free and correlated
neutrons. Here, we will suppose that the quasiparticle dispersion
relation $\epsilon_{\vekk}$ is, up to a possible constant shift that
can be absorbed in the chemical potential, close to the free one,
$\epsilon_{\vekk}\approx k^2/2$. The effect of the modification of the
momentum dependence of the quasiparticle dispersion (effective mass)
on the BCS gap, calculated within the Hartree-Fock (HF) approximation
using chiral low-momentum interactions including three-body force, was
studied in~\cite{HebelerSchwenk2010} and it was found that this effect
is very weak at low densities (see Fig.~10 of
\cite{HebelerSchwenk2010}).

\begin{figure}
\includegraphics[angle = 0, width = 3.4in, clip = true]{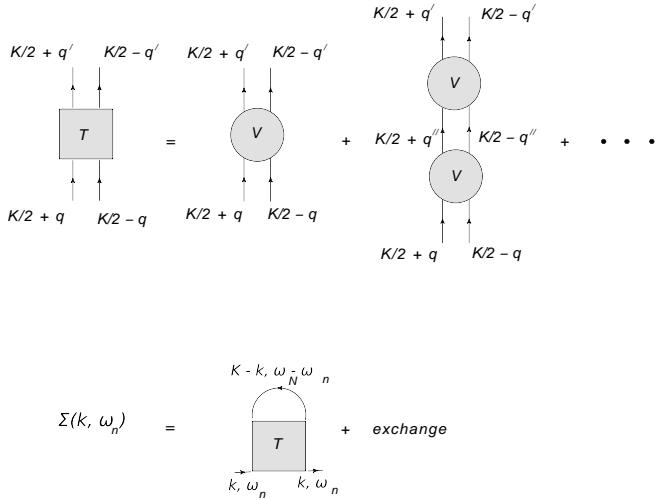}
\caption{Feynman Diagrams for the $T$ matrix and the self energy.}
\label{fig:feyn}
\end{figure}

Within the NSR approach, the density of neutrons at finite
temperatures can be decomposed into the following pieces:
\begin{equation} \rho = \rhofree + \rhocorr.
\label{eq:NSR_decomp}
\end{equation}
The first term, $\rhofree$, is the density of uncorrelated neutrons,
\begin{equation}
\rhofree = 2\int \frac{d^3k}{(2\pi)^3} f(\xi_{\vekk})\,,
\label{eq:rho_free}
\end{equation}
where the factor of $2$ accounts for the spin degeneracy. The second
term, $\rhocorr$, is the contribution due to two neutron scattering
and is given by
\begin{equation} 
\rhocorr = 2 \int \frac{d^3k}{(2 \pi)^3} 
  \frac{1}{\beta} \sum_{\omega_n} \left(\Gtemp_0^{(1)}(\vekk, \omega_n)\right)^2 
  \Sigma(\vekk, \omega_n),
\label{eq:rhocorr}
\end{equation}
The single-particle self-energy $\Sigma$ (see Fig.~\ref{fig:feyn})
%%%%%%%%%%%%%%%%%%%%%%%%%%%%%%%%%%%%%%%%%%%%%%%%%%%%%%%%%%%%%%%%%%%%%%%%%%%%
%%%%%%%%%%%%%%%%%%%%%%%%%%%%%%%%%%%%%%%%%%%%%%%%%%%%%%%%%%%%%%%%%%%%%%%%%%%%
is calculated in ladder approximation, i.e.,
\begin{multline}
\Sigma(\vekk, \omega_n) = \int \frac{d^3K}{(2\pi)^3}
  \frac{1}{\beta} \sum_{\omega_N}
  \Gtemp_0^{(1)}(\vekK - \vekk, \omega_N - \omega_n)\\ 
  \times T\Big(\vekK, \frac{\vekK}{2} - \vekk, 
    \frac{\vekK}{2} - \vekk, \omega_N\Big),
\label{eq:self_en1}
\end{multline}
with the neutron-neutron $T$ matrix that satisfies the
following equation:
\begin{multline}
T(\vekK, \vekq, \vekq', \omega_N)
  = V(\vekq, \vekq')\\
  + \int \frac{d^3 q''}{(2\pi)^3} 
    V(\vekq, \vekq'') \Gtemp_0^{(2)}(\vekK, \vekq'', \omega_N)
    V(\vekq'',\vekq')\\
  +\cdots\,.
\label{eq:ladder}
\end{multline}

Substituting Eq.~(\ref{eq:self_en1}) into Eq.~(\ref{eq:rhocorr}) and using
$(\Gtemp_0^{(1)})^2 = -\partial\Gtemp_0^{(1)}/\partial\mu$, we get
\begin{widetext}
\begin{equation}
\rhocorr = -2 \int \frac{d^3 K}{(2\pi)^3} 
  \frac{1}{\beta} \sum_{\omega_N}
  \int \frac{d^3q}{(2\pi)^3} 
  \frac{1}{\beta} \sum_{\omega_n}
  \left(\frac{\partial}{\partial \mu}
     \Gtemp_0^{(1)}\Big(\frac{\vekK}{2} + \vekq, \omega_n\Big)\right)
     \Gtemp_0^{(1)}\Big(\frac{\vekK}{2} - \vekq, \omega_N - \omega_n\Big) 
     T(\vekK, -\vekq, -\vekq, \omega_N).
\label{eq:rhocorr2}
\end{equation}
\end{widetext}
Next we expand the $T$ matrix in a partial-wave basis and pick out the
$s$-wave ($l=0$) contribution:
\begin{equation}
  T(\vekK, -\vekq, -\vekq, \omega_N) = 4 \pi T_{l=0}(K, q, q, \omega_N).
\label{eq:tmat_even_odd}
\end{equation}
Since we can replace $\vekq$ by $-\vekq$ in Eq.~(\ref{eq:rhocorr2}), the
derivative $\partial/\partial\mu$ can act on either of the Green's
functions $\Gtemp_0^{(1)}$ and we can therefore let it act on both if we
multiply by a factor of $\frac{1}{2}$. Using the definition
(\ref{eq:two_body_gf}), we may therefore write
\begin{multline}
\rhocorr = \int \frac{d^3K}{(2 \pi)^3} \frac{1}{\beta} \sum_{\omega_N} 
  \int\frac{d^3q}{(2 \pi)^3} \left(\frac{\partial}{\partial \mu} 
  \Gtemp_0^{(2)}(\vekK,\vekq, \omega_N)\right)\\
  \times T_{l=0}(K, q, q, \omega_N)\,.
\label{eq:rhocorr3}
\end{multline}
Now we rewrite also Eq.~(\ref{eq:ladder}) in a partial-wave basis and
insert it into Eq.~(\ref{eq:rhocorr3}). Since we consider only the $s$
wave, we may replace $\Gtemp_0^{(2)}$ by its angle average
$\avGtemp_0^{(2)}$. In an analogous way as explained above, we account
for the derivative $\partial/\partial\mu$ acting on all the
$\avGtemp_0^{(2)}$ if we multiply the $i$-th term by a factor
$\frac{1}{i}$:
\begin{multline}
\rhocorr = \frac{\partial}{\partial \mu} \int \frac{K^2 dK}{2 \pi^2}
   \frac{1}{\beta} \sum_{\omega_N}\Big[
   \frac{2}{\pi} \int q^2 dq \,\avGtemp_0^{(2)}(q) v(q,q)\\
    + \frac{1}{2}\, \frac{2}{\pi} \int q^2 dq 
      \frac{2}{\pi} \int q'^2 dq'
      \avGtemp_0^{(2)}(q) v(q,q')
      \avGtemp_0^{(2)}(q') v(q',q)\\
    + \cdots \Big]\,.
\label{eq:rhocorr4}
\end{multline}
For brevity, the arguments $K$ and $\omega_N$ have been dropped in
$\avGtemp_0^{(2)}$. The terms in the square brackets can be summed so
that
\begin{equation}
\rhocorr = -\frac{\partial}{\partial \mu} \int \frac{K^2 dK}{2 \pi^2}
  \frac{1}{\beta} \sum_{\omega_N} 
  \tr\,\log\left(1 - \avGtemp_0^{(2)} v\right)\,,
\label{eq:rhocorr5}
\end{equation}
where the trace corresponds to the integral over $q$, the product
$\avGtemp_0^{(2)} v$ has to be understood as a product of matrices having
continuous indices $q$, $q'$, and the logarithm is a matrix
logarithm defined by its power series.

Using an analytic continuation as in
Eq.~(\ref{eq:two_body_gf_retarded}) and standard techniques to
transform the frequency sum into an integral over real energies
\cite{Fetter_walecka}, we obtain
\begin{multline}
\rhocorr = -\frac{\partial}{\partial \mu} \int \frac{K^2 dK}{2 \pi^2}
  \int \frac{d\omega}{\pi}\, g(\omega)\\
  \times \im\tr\log\left(1 - \avG_0 v\right)\,,
\label{eq:rhocorr_real}
\end{multline}
where $\avG_0$ is the retarded two-body Green's function given
in Eq.~(\ref{eq:avG0}) and $g(\omega) = 1/(e^{\beta\omega}-1)$ is the
Bose function. Note that in spite of the pole of the Bose function the
integrand is well behaved at $\omega = 0$ because $\avQ(K,q)$
vanishes at $K^2/4+q^2=2\mu$ and therefore $\im \avG_0$ vanishes
at $\omega=0$. Since the trace is invariant under a change of the
basis, it is most convenient to calculate it in the basis of Weinberg
eigenvectors where $\avG_0 v$ is diagonal. In this way we get
\begin{multline}
\rhocorr = -\frac{\partial}{\partial \mu} \int \frac{K^2 dK}{2 \pi^2}
  \int \frac{d\omega}{\pi}\, g(\omega)\\ 
  \times \im \sum_\nu \log\big(1 - \eta_\nu(K,\omega)\big)\,.
\label{eq:rhocorr_eta}
\end{multline}

The expression (\ref{eq:rhocorr}) for $\rhocorr$ corresponding to the
original NSR scheme \cite{NSR} does not only describe the effect of
correlations. A non-vanishing real part of the on-shell self-energy
shifts the quasiparticle energies and thereby strongly affects the
density at fixed chemical potential. However, there is no reason to
assume that the shift calculated with only the two-body interaction in
the $^1S_0$ channel is realistic, and as mentioned before, this shift
should be already included in the quasiparticle dispersion relation
$\epsilon_{\vekk}$. One should therefore subtract the real part
  of the on-shell self-energy $\Sigma^R(\vekk,\xi_{\vekk})$
[where $\Sigma^R$ denotes the retarded self-energy, related to
    the imaginary-time self-energy $\Sigma$ of Eq.~(\ref{eq:self_en1})
    by analytic continuation] from the self-energy in
Eq.~(\ref{eq:rhocorr}). This prescription was used in symmetric matter
\cite{rostock1,rostock2,Jin:2010nj}, see appendix for more
  details. In the present work, we will use a slightly simplified
prescription, namely we will subtract only the energy-independent
leading term of the self-energy,
\begin{equation}
\Sigma_1(\vekk) = \int \frac{d^3p}{(2 \pi)^3} 
  V\Big(\frac{\vekk - \vekp}{2},\frac{\vekk - \vekp}{2}\Big) f(\xi_{\vekp})\,.
\label{eq:sub_sf}
\end{equation}
This corresponds to the HF potential, except that it is not
calculated self-consistently. The change in density due to $\Sigma_1$
is given by
\begin{equation}
\rho_1 = 2\int\frac{d^3k}{(2\pi)^3}
  \frac{\partial f(\xi_{\vekk})}{\partial\xi_{\vekk}} \Sigma_1(\vekk)\,.
\end{equation}
With the help of the property $f(\xi_{\vekk})f(\xi_{\vekp}) =
g(\xi_{\vekk}+\xi_{\vekp})[1-f(\xi_{\vekk})-f(\xi_{\vekp})]$ and after
transformation to total and relative momenta, this can be rewritten in
a form with no angular integrals as
\begin{multline}	
\rho_1 = -\frac{\partial}{\partial \mu} \int
\frac{K^2 dK}{2\pi^2} \frac{2}{\pi}\int q^2 dq\, 
  g\Big(\frac{K^2}{4}+q^2-2\mu\Big) \\
  \times v(q,q)\,\avQ(K, q)\,,
\label{eq:sub_den}
\end{multline}
The corrected correlated density is now given by
$\rhocorr-\rho_1$.

We have seen that the use of the Weinberg eigenvector basis, where
$\avG_0 v$ is diagonal, offers a convenient way to apply the NSR
scheme to the case of a non-local and non-separable interaction, such
as the low-momentum interaction $\vlowk$. We can now study the total
density at different temperatures above the transition temperature and
analyze the importance of correlations. We present our results in the
next section.

%%%%%%%%%%%%%%%%%%%%%%%%%%%%%%%%%%%%%%%%%%%%%%%%%%%%%%%%%%%%%%%%%%%%%
\section{Results and Discussion}
\label{sect:results}
%%%%%%%%%%%%%%%%%%%%%%%%%%%%%%%%%%%%%%%%%%%%%%%%%%%%%%%%%%%%%%%%%%%%%

\begin{figure}[t]
\begin{center}
\includegraphics[angle=0, width=7.75cm,
  clip=true]{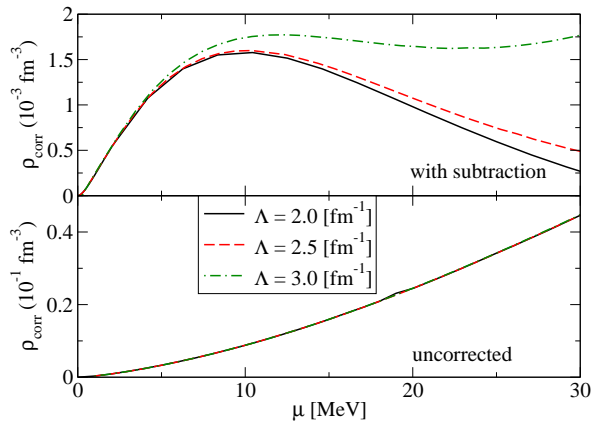}
  \end{center}
  \caption{(color online) Correlated density $\rhocorr$ at $T =
    T_c(\mu)$ with (upper panel) and without (lower panel) the
    correction (\ref{eq:sub_den}) as functions of the chemical
    potential $\mu$. The different curves were obtained with $\vlowk$
    interactions corresponding to cutoffs $\Lambda = 2,$ $2.5$, and $3
    \fmi$.}
  \label{fig:rho-corr_lam}
 \end{figure}

In section~\ref{sect:T_c} we presented a method to obtain the
transition temperature $T_c$ as a function of the chemical potential
$\mu$ using the Weinberg eigenvalues. In BCS theory, one assumes that
there are no correlations above $T_c$, and therefore one computes the
transition temperature as a function of the density $\rho$ by using
$\rho = \rhofree$ according to Eq.~(\ref{eq:rho_free}). However, as
discussed in the preceding section, the presence of correlated pairs
above $T_c$ changes the relation between $\mu$ and $\rho$. In
Fig.~\ref{fig:rho-corr_lam},
%%%%%%%%%%%%%%%%%%%%%%%%%%%%%%%%%%%%%%%%%%%%%%%%%%%%%%%%%%%%%%%%%%%%%%%%%
 %%%%%%%%%%%%%%%%%%%%%%%%%%%%%%%%%%%%%%%%%%%%%%%%%%%%%%%%%%%%%%%%%%%%%%%%%
we show our results for the correlated density $\rhocorr$ at the
transition temperature $T_c$ with (upper panel) and without (lower
panel) the subtraction of the contribution due to the HF like shift,
Eq.~(\ref{eq:sub_den}), as functions of the chemical potential
$\mu$. We observe that without the subtraction, $\rhocorr$ is
monotonically increasing, while the subtraction reduces it a lot,
especially at large $\mu$ (high density). We did the calculations for
different values of the cutoff $\Lambda$ between $2$ and $3\fmi$, and
as it was the case for $T_c$ (cf.\ Fig.~\ref{fig:trans_T_vs_mu}), the
cutoff dependence of $\rhocorr$ without the correction for the HF like
shift is very weak. Therefore, one can conclude that the correlated
density \emph{without} the subtraction as defined in~\ref{sect:NSR} is
well constrained by the NN phase shift analogous to the transition
temperature $T_c$. Such weak cut-off dependence was already seen in
the BCS pairing gaps in~\cite{Hebeler2007}.  However, we note that the
\emph{corrected} densities depend on the cutoff, especially at high
density. This dependence can be traced back to the HF approximation of
the subtracted on-shell self energy. Only with low cutoffs where the
interaction has been sufficiently softened, will the HF self-energy
subtraction suffice, while in the case of $\Lambda = 3\fmi$ the
correlated density $\rhocorr$ does not vanish at large $\mu$ (i.e.,
high density). The cut-off dependence of the HF contribution at high
density is an indication for missing three-body and other higher order
effects.

\begin{figure}[t]
 \begin{center}
\includegraphics[angle=0, width=7.5cm,
  clip=true]{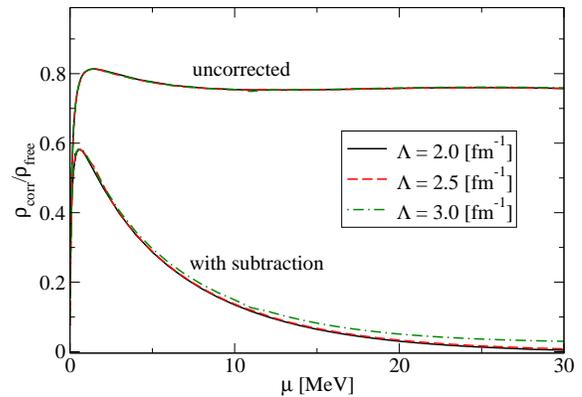}
\end{center}
\caption{Ratio of the correlated density to the free density
  ($\rhocorr/\rhofree = \rho/\rhofree-1$) as function of the chemical
  potential $\mu$ at $T=T_c(\mu)$ with (lower curves) and without
    (upper curve) the subtraction term (\ref{eq:sub_den}). The
    different curves correspond to cut-offs $\Lambda = 2$, $2.5$, and
    $3 \fmi$. We see that the cut-off dependence is drastically
    lowered as compared to the upper panel of
    Fig.~\ref{fig:rho-corr_lam} because of the dominant contribution
    from the free density.}
\label{fig:ratios_vs_mu}
\end{figure}

Fortunately, at the densities where the cutoff dependence from this
correction is significant, the total density is completely dominated
by the free one. Therefore, the total density is only weakly cutoff
dependent. This can be better understood from
Fig.~\ref{fig:ratios_vs_mu},
%%%%%%%%%%%%%%%%%%%%%%%%%%%%%%%%%%%%%%%%%%%%%%%%%%%%%%%%%%%%%%%%%%%%%%%%%
%%%%%%%%%%%%%%%%%%%%%%%%%%%%%%%%%%%%%%%%%%%%%%%%%%%%%%%%%%%%%%%%%%%%%%%%%
where we show the ratio $\rhocorr/\rhofree$ of the correlated density
at $T = T_c$ to the free density as a function of the chemical
potential $\mu$, again for cutoffs $\Lambda = 2$, $2.5$, and $3
  \fmi$. The figure also shows the effect of the HF like
  subtraction. The lower curves are the ratios calculated with the
  subtraction (\ref{eq:sub_den}), while the upper curve does not have
the correction. Without the subtraction, we can clearly see the
over-all change of the densities due to the HF like shift of the
quasi-particle energies. Once we include the subtraction, the
densities converge towards the free ones in the limit of large $\mu$,
i.e., at high densities. This shows clearly that at high densities the
shift comes only from the first-order (HF like) term and has nothing
to do with correlated pairs. This was expected, since at high
densities, pair correlations are weak, as one can see from the low
critical temperature in Fig.~\ref{fig:trans_T_vs_mu}. However, at low
densities, the correlations lead to a sizeable enhancement of the
density. In other words, if one fixes the density $\rho$, the
correlations lead to a reduction of the chemical potential $\mu$. This
results in a reduction of the critical temperature $T_c$ as a function
of $\rho$ with respect to the BCS one.

\begin{figure}
\begin{center}
\includegraphics[angle = 0, width = 7.5cm, clip = true]
  {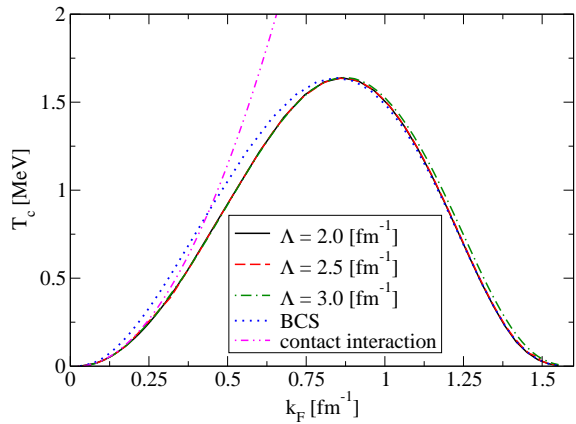}
\end{center}
\caption{(color online) Transition temperature $T_c$ as a function of
  $k_F = (3\pi^2\rho)^{1/3}$. Solid, dashed, and dashed-dotted
    lines: full NSR results with correction obtained with $\vlowk$
    interactions corresponding to cutoffs $\Lambda = 2$, $2.5$ and $3
    \fmi$; dotted line: corresponding BCS results ($\rho =
  \rhofree$); dashed-double dotted line: result for a contact
  interaction corresponding to a scattering length $a=-18\fm$ within
  the original NSR scheme \cite{SadeMelo1993}.}
\label{fig:tc_vs_rho}
\end{figure}

This brings us to the main result of our study, namely the density
dependence of the transition temperature, displayed in
Fig.~\ref{fig:tc_vs_rho}.
%%%%%%%%%%%%%%%%%%%%%%%%%%%%%%%%%%%%%%%%%%%%%%%%%%%%%%%%%%%%%%%%%%%%%%%%%
%%%%%%%%%%%%%%%%%%%%%%%%%%%%%%%%%%%%%%%%%%%%%%%%%%%%%%%%%%%%%%%%%%%%%%%%%
The BCS result, obtained with $\rho(\mu) = \rhofree(\mu)$, is shown as
the dotted line. If we calculate the total density according to
Eq.~(\ref{eq:NSR_decomp}), including the correction in
Eq.~(\ref{eq:sub_den}), we obtain the NSR results shown as the solid,
dashed, and dashed-dotted lines (corresponding again to the three
  cutoffs $\Lambda=2$, $2.5$, and $3 \fmi$) We note that the BCS and
NSR results agree above $k_F\sim 0.8\fmi$, i.e., at densities above
$~0.017\fmithree\sim 0.1\rho_0$ ($\rho_0 = 0.17\fmithree$ being the
saturation density of nuclear matter), as one could have anticipated
from the vanishing of $\rhocorr$ at high density. Notice that within
the original NSR scheme, i.e., without the subtraction of
Eq.~(\ref{eq:sub_den}), this would not have been the case. At lower
densities, the NSR transition temperature is significantly lower than
the BCS one. Not surprisingly, the pair correlations above $T_c$ are
most important at low densities, where the neutron gas is in the
BCS-BEC crossover regime close to the unitary limit.

For comparison we show as the dashed-double dotted line the results
obtained within the NSR scheme with a contact interaction
\cite{SadeMelo1993}. The contact interaction is characterized by the
scattering length that is set to $a=-18\fm$. By fixing the scattering
length, the coupling constant and the cutoff are related, and when one
takes the cutoff to infinity the coupling constant goes to
zero. Therefore there is no correction from the subtraction of the
first-order (HF like) contribution in this scheme. We see that the
results of the full calculation are in good agreement with the results
for the contact interaction up to $k_F \sim 0.2\fmi$, i.e., only up to
a tiny density of $\sim 0.0003 \fmithree \sim 0.002 \rho_0$. At higher
densities, the combination of the finite range of the interaction and
Pauli blocking leads to a suppression of pairing correlations.

By comparing the results obtained with different cutoffs (solid,
  dashed, and dashed-dotted curves in Fig.~\ref{fig:tc_vs_rho}), one
  observes a weak cut-off dependence that arises at high
  densities. This sets the scale for the missing higher-order
contributions as well the missing three and higher-body
forces. Analogous to Fig.~\ref{fig:rho-corr_lam}, we see that the
cut-off dependence is weak as it is over-whelmed by the contributions
from $\rho_{\text{free}}$ at high densities due to Pauli blocking. The
cut-off dependence seen here will be different if one were to, for
example, include corrections to the quasi-particle energies. These
effects become especially important at high
densities~\cite{HebelerSchwenk2010}.

\begin{figure}
\begin{center}
\includegraphics[angle=0, width=8.5cm, clip=true]{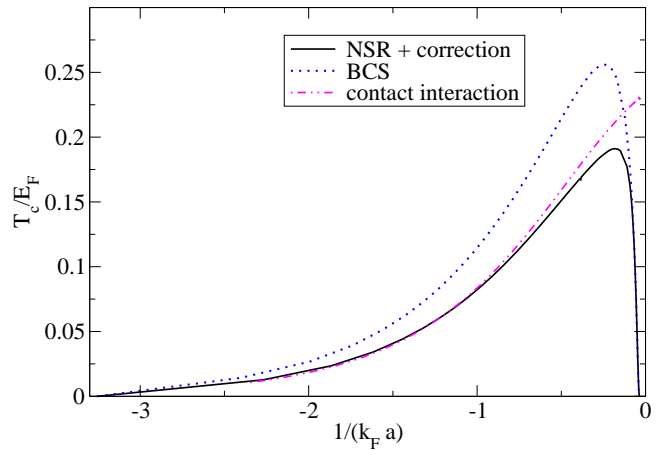}
\end{center}
\caption{(color online) Same results as shown in
  Fig.~\ref{fig:tc_vs_rho}, but $T_c$ is scaled by the Fermi energy
  $E_F = k_F^2/2$ and shown as function of the dimensionless BEC-BCS
  crossover parameter $1/(k_Fa)$.}
\label{fig:tcef_vs_kfa}
\end{figure}

Finally, in order to discuss the connection between low-density
neutron matter and the BCS-BEC crossover, we show in
Fig.~\ref{fig:tcef_vs_kfa}
%%%%%%%%%%%%%%%%%%%%%%%%%%%%%%%%%%%%%%%%%%%%%%%%%%%%%%%%%%%%%%%%%%%%%%%%%
%%%%%%%%%%%%%%%%%%%%%%%%%%%%%%%%%%%%%%%%%%%%%%%%%%%%%%%%%%%%%%%%%%%%%%%%%
the same results as in Fig.~\ref{fig:tc_vs_rho}, but plotted in a
different way. In the case of a contact interaction, the crossover is
characterized by the dimensionless parameter $1/(k_Fa)$: for
$1/(k_Fa)\ll -1$, the system is in the BCS regime; for $|1/(k_Fa)|
\lesssim 1$, the system is in the crossover regime, $1/(k_Fa)=0$
corresponding to the unitary limit; and for $1/(k_Fa)\gg 1$, the
system forms a BEC of bound dimers. While in experiments with
ultracold atoms the whole crossover can be studied by varying $a$
\cite{Regal2004,Zwierlein2004}, only the region of negative $1/(k_Fa)$
is accessible in neutron matter by varying $k_F$. In
Fig.~\ref{fig:tcef_vs_kfa} we show the dimensionless ratio $T_c/E_F$,
($E_F$ being the Fermi energy) in neutron matter within NSR (solid
line) and BCS (dotted line) as well as the result for a contact
interaction with $a = -18\fm$ within the NSR approach
(dashed-double dotted line), as functions of $1/(k_Fa)$. As we
already observed in Fig.~\ref{fig:tc_vs_rho}, the NSR results for the
neutron-neutron and contact interactions are in good agreement at very
low densities [$1/(k_Fa)\lesssim -0.8$], while at $1/(k_Fa) \gtrsim
-0.1$ the finite range of the neutron-neutron interaction leads to a
strong suppression of pairing correlations so that the system returns
to the BCS regime. The point where neutron matter is closest to the
unitary Fermi gas is $1/(k_Fa) \sim -0.2$, corresponding to a very
small density of $\sim 0.0007\fmithree \sim 0.004\rho_0$.

On a quantitative level, there still remain strong
uncertainties. Although we find at low densities a significant
reduction of $T_c$ with respect to the BCS result, other many-body
effects may lead to additional suppression of $T_c$. For instance, the
inclusion of the HF quasiparticle effective mass, $m^* < m$, reduces
the density of states and therefore the critical temperature. In this
case, also the cutoff-independence of the results would be lost. Such
effects were studied, e.g., in \cite{HebelerSchwenk2010}, and they are
important at higher density, but not in the cross-over region on which
we are focusing here. Screening of the interaction by the medium,
which is not included in the present calculation either, can also
result in a dramatic suppression of pairing~\cite{Shen2002}. That
corrections beyond NSR have to play a role can be seen, for instance,
in the case of the contact interaction in the unitary limit ($a\to
\infty$). In this case, the NSR scheme predicts a critical temperature
of $T_c/E_F=0.23$ \cite{SadeMelo1993}, which is strongly reduced
compared with the corresponding BCS result of $T_c = 0.49 E_F$, but
still higher than the experimental value $T_c = 0.167 E_F$ measured in
an ultracold gas of trapped $^6$Li atoms~\cite{Ku2012}. These numbers
suggest that, in the case of a contact interaction near the unitary
limit, the NSR scheme is able to describe the essential
effect. However, it is possible that in neutron matter, where the
interaction is much more complex than in ultracold atoms, screening
corrections might be more important. This question needs further
investigations.

%%%%%%%%%%%%%%%%%%%%%%%%%%%%%%%%%%%%%%%%%%%%%%%%%%%%%%%

\section{Summary and Outlook}
\label{sect:summary}

In this paper, we study the effect of correlations above the
superfluid transition temperature in neutron matter in the BEC-BCS
crossover regime within the Nozi\`eres-Schmitt-Rink scheme. We use as
input the renormalization group based low-momentum effective
interaction $\vlowk$. In order to deal with the non-local interaction,
we use the Weinberg basis, where the operator $G_0 V$ is diagonal. Our
results show that the transition temperature is lower than the BCS
result at low densities, while at high densities we get back the BCS
result. At very low densities, our results are in reasonable agreement
with those obtained with a contact interaction.

Our main goal is to demonstrate the importance of beyond BCS physics
in neutron matter in the crossover regime, and our study is far from
being exhaustive. For instance, we use a free particle spectrum and
correct for the constant shift of quasi-particle energies using the
static Hartree-Fock approximation. We also neglect three-body forces
for the sake of simplicity.

It would certainly be interesting to incorporate a three-body force or
at least a density dependent two-body force, although one expects such
effects to be important at higher densities~\cite{HebelerSchwenk2010}
where the BCS results hold. A calculation including the three-body
force would be far more involved. For example, if one includes a
density dependent two-body force, the interaction will have to be
evaluated self-consistently as the correlations are built into the
total density. Similarly, a subtraction of the full on-shell
self-energy within the ladder approximation would reduce the cut-off
dependence seen in
Figs.~\ref{fig:rho-corr_lam},~\ref{fig:ratios_vs_mu}
and~\ref{fig:tc_vs_rho}, but such a calculation is beyond the scope of
the current work, although it is worth-while investigating in
future. 

Finally, other many-body effects like particle-hole correlations
  (screening) \cite{Shen2002} are also important. In the low-density
  limit, these effects are known to reduce the critical temperature by
  a factor of $1/(4e)^{1/3} \approx 0.45$ (Gor'kov-Melik-Barkhudarov
  correction \cite{GorkovMelikBarkhudarov}). Therefore, they should be
  included in future studies.

%%%%%%%%%%%%%%%%% Acknowledgements %%%%%%%%%%%%%%%%%%%%%%%%%%%%%%%%%%
\begin{acknowledgments}
We would like to thank Peter Schuck for discussions and Dick Furnstahl
for comments on the manuscript. SR acknowledges the support and
hospitality of IPN Orsay where a major part of the work was carried
out. SR also acknowledges the support from the HPC cluster at IIT
Madras, Chennai, India.
\end{acknowledgments}

%%%%%%%%%%%%%%%%%%%%%%%%% Appendix %%%%%%%%%%%%%%%%%%%%%%%%%%%%%%%%%%
\appendix*
\section{Justification of the subtraction}
In this appendix we wish to explain in more detail the necessity of
subtracting the mean-field shift from the self-energy when calculating
the correlated density. As pointed out in section~\ref{sect:NSR}, a
problem of the original NSR approach \cite{NSR} is that the
self-energy, including its energy-independent part leading to a
mean-field-like shift of the single-particle energies, is treated only
perturbatively. But the shift of the single-particle energies results
in a strong correction of the density which has nothing to do with
pair correlations and which in a more consistent calculation would be
absorbed to a large extent in a corrected chemical potential.

In the NSR approach, the self-energy $\Sigma$ is calculated with free
propagators $\Gtemp_0^{(1)}$ given by Eq.~(\ref{eq:sing_gf}). However,
in a more complete calculation, one should use (self-consistent)
dressed Green's functions
\begin{equation}
\Gtemp^{(1)}(k,\omega_n) =
\frac{1}{i\omega_n-\xi_k-\Sigma(k,\omega_n)}
\end{equation}
throughout the calculation of $\Sigma$. Since this is very
difficult, one should at least approximate the dressed Green's
functions by quasiparticle ones
\begin{equation}
\tilde{\Gtemp}_0^{(1)}(k,\omega_n) =
\frac{1}{i\omega_n-\tilde{\xi}_k}
\end{equation}
where the quasiparticle energy $\tilde{\xi}_k$ includes the
shift due to the real part of the self-energy and is determined
self-consistently as the solution of
\begin{equation}
\tilde{\xi}_k = \xi_k + \re \Sigma^R(k,\tilde{\xi}_k)\,.
\end{equation}
Applying the BCS approximation within this quasiparticle picture, the
critical temperature $T_c$ is determined from the pole in the $T$
matrix, calculated with $\tilde{\Gtemp}_0^{(1)}$ instead of
$\Gtemp_0^{(1)}$, and the density $\rho$ from Eq.~(\ref{eq:rho_free})
with $\tilde{\xi}_k$ instead of $\xi_k$. The main effect of replacing
particles by quasiparticles comes from the effective mass [$m^* =
  k_F/(d\tilde{\xi}_k/{dk})_{k=k_F}$], since it changes the density of
states, while a momentum-independent shift of the single-particle
energies has no effect at all on the relation between $T_c$ and $\rho$
since it can be absorbed in an effective chemical potential $\mu^*$.

Let us now go beyond the BCS approximation. In the approach developed
in Ref.~\cite{ZimmermannStolz} in the context of solid-state physics
and applied to nuclear matter in
Refs.~\cite{rostock1,rostock2,Jin:2010nj}, only the correlation
contribution, i.e., the energy-dependent part of the self-energy, is
treated perturbatively. Then the approximation for the corrected
single-particle Green's function reads
\begin{multline}
\Gtemp^{(1)}(k, \omega_n) = \tilde{\Gtemp}_0^{(1)}(k, \omega_n) \\
  + \left(\tilde{\Gtemp}_0^{(1)}(k, \omega_n)\right)^2 
  \left(\Sigma(k, \omega_n)-\re \Sigma^R(k,\tilde{\xi}_k)\right)\,.
\end{multline}
Note that $\re \Sigma^R(k,\tilde{\xi}_k)$ has to be subtracted from
$\Sigma(k, \omega_n)$ since it is already contained in
$\tilde{\Gtemp}_0^{(1)}$. The density is now obtained by summing
$\Gtemp^{(1)}(k, \omega_n)$ over $\omega_n$ and integrating over
$\vekk$, which gives an equation for $\rhocorr$ analogous to
Eq.~(\ref{eq:rhocorr}) but with a subtraction term.

To arrive at the subtraction we use in section~\ref{sect:NSR}, two
additional approximations are made. First, we assume that the
single-particle spectrum $\tilde{\xi}_k$ can be approximated by
$\tilde{\xi}_k \approx \epsilon_k-\mu^*$, i.e., we neglect the
effective mass and other more complicated momentum dependences of $\re
\Sigma^R(k,\tilde{\xi}_k)$. As it was shown in
\cite{HebelerSchwenk2010} (where $\tilde{\xi}_k$ was calculated in the
HF approximation), these effects are not important in the low-density
region we focus on. Note that now the quantity called $\mu$ in
sections~\ref{sect:T_c}-\ref{sect:results} is not the real
chemical potential but the effective one, $\mu^*$, which includes the
mean-field shift.

Second, we replace $\re \Sigma^R(k,\tilde{\xi}_k)$ in the subtraction
term by the first-order (HF) term $\Sigma_1(k)$,
Eq.~(\ref{eq:sub_sf}). The reason for this is a purely practical
one. While in the case of a separable potential the correlated density
with subtraction can be reduced to a simple expression containing only
the in-medium scattering phase shift
\cite{rostock1,rostock2,Jin:2010nj,ZimmermannStolz}, we did not
succeed to derive an analogous formula in the case of our
non-separable interaction. The direct calculation of the on-shell
self-energy $\Sigma^R(k,\tilde{\xi}_k)$, however, is numerically quite
involved. This approximation to replace $\re
\Sigma^R(k,\tilde{\xi}_k)$ by $\Sigma_1(k)$ is not valid for large
cut-offs and this is the main reason for the cutoff dependence of our
results with subtraction.
%%%%%%%%%%%%%%%%% References %%%%%%%%%%%%%%%%%%%%%%%%%%%%%%%%%%%%%%


\begin{thebibliography}{99} 

\bibitem{Margueron:2007uk} 
  J.~Margueron, H.~Sagawa and K.~Hagino,
  %``BCS-BEC crossover of neutron pairs in symmetric and asymmetric
  %nuclear matters,''
  Phys.\ Rev.\ C {\bf 76}, 064316 (2007).
  
\bibitem{Baldo:2004dx} 
  M.~Baldo, C.~Maieron, P.~Schuck and X.~Vinas,
  %``Low densities in nuclear and neutron matters and in the nuclear
  %surface,''
  Nucl.\ Phys.\ A {\bf 736}, 241 (2004).
  
\bibitem{Monrozeau:2007xu} 
  C.~Monrozeau, J.~Margueron and N.~Sandulescu,
  %``Nuclear superfluidity and cooling time of neutron-star crust,''
  Phys.\ Rev.\ C {\bf 75}, 065807 (2007).

\bibitem{Yakovlev:2004iq} 
  D.~G.~Yakovlev and C.~J.~Pethick,
  %``Neutron star cooling,''
  Ann.\ Rev.\ Astron.\ Astrophys.\  {\bf 42}, 169 (2004).

\bibitem{Hagino:2006ib} 
  K.~Hagino, H.~Sagawa, J.~Carbonell and P.~Schuck,
  %``Coexistence of BCS and BEC-like pair structures in halo nuclei,''
  Phys.\ Rev.\ Lett.\  {\bf 99}, 022506 (2007).
  
\bibitem{Sagawa:2012je} 
  H.~Sagawa and K.~Hagino,
  %``Pairing correlations in exotic nuclei,''
  J.\ Phys.\ Conf.\ Ser.\  {\bf 413}, 012008 (2013).
  
\bibitem{Sagawa:2007we} 
  H.~Sagawa and K.~Hagino,
  %``Pairing correlations in halo nuclei,''
  e-print arXiv:0709.1310 [nucl-ex] (2007).

\bibitem{Hagino:2005we} 
  K.~Hagino and H.~Sagawa,
  %``Pairing correlations in nuclei on the neutron-drip line,''
  Phys.\ Rev.\ C {\bf 72}, 044321 (2005).
  
\bibitem{rostock1} M.~Schmidt, G.~R\"opke, and H.~Schulz,
  Ann.\ Phys.\ (N.Y.) {\bf 202}, 57 (1990).

\bibitem{rostock2} H.~Stein, A.~Schnell, T.~Alm, and
  G.~R\"opke, Z.\ Phys.\ A {\bf 351}, 295 (1995).

\bibitem{Jin:2010nj} 
  M.~Jin, M.~Urban and P.~Schuck,
  %``BEC-BCS Crossover and the Liquid-Gas Phase Transition in Hot and Dense
  %Nuclear Matter,''
  Phys.\ Rev.\ C {\bf 82}, 024911 (2010).

\bibitem{Baker1999} G.~A.~Baker Jr., Phys.\ Rev.\ C {\bf 60}, 054311
  (1999).

\bibitem{Gonzalez1999} D.~E.~Gonz\'alez~Trotter {\it et al.},
  Phys.\ Rev.\ Lett. {\bf 83}, 3788 (1999).

\bibitem{ChamelHaensel2008} N.~Chamel and P.~Haensel, Living
  Rev.\ Relativity {\bf 11}, 10 (2008)
  [http://www.livingreviews.org/lrr-2008-10].

\bibitem{NSR} P.~Nozi\`eres and S.~Schmitt-Rink, J.\ Low
  Temp.\ Phys.\ {\bf 59}, 195, (1985).

\bibitem{Bogner_rev2009} 
  S.~K.~Bogner, R.~J.~Furnstahl and A.~Schwenk,
  %``From low-momentum interactions to nuclear structure,''
  Prog.\ Part.\ Nucl.\ Phys.\  {\bf 65}, 94 (2010)
  [arXiv:0912.3688 [nucl-th]].
  %%CITATION = ARXIV:0912.3688;%%
  %112 citations counted in INSPIRE as of 28 Jul 2013
  
\bibitem{Hebeler2007} K.~Hebeler, A.~Schwenk, and B.~Friman,
  Phys.\ Lett.\ B {\bf 648}, 176 (2007).

\bibitem{Lesinski2009} T.~Lesinski, T.~Duguet, K.~Bennaceur, and
  J.~Meyer, Eur.\ Phys.\ J.\ A {\bf 40}, 121 (2009)

\bibitem{Hebeler2009} K.~Hebeler, T.~Duguet, T.~Lesinski, and
  A.~Schwenk, Phys.\ Rev.\ C {\bf 80}, 044321 (2009).

\bibitem{Weinberg} S.~Weinberg, Phys.\ Rev.\ {\bf 131} 440 (1963).

\bibitem{bogner2006}
  S.~K.~Bogner, R.~J.~Furnstahl, S.~Ramanan, and A.~Schwenk,
  % ``Convergence of the Born Series with Low-Momentum Interactions,''
  %
  Nucl.\ Phys.\  {\bf A773} (2006) 203.

\bibitem{Bogner:2005sn} 
  S.~K.~Bogner, A.~Schwenk, R.~J.~Furnstahl and A.~Nogga,
  %``Is nuclear matter perturbative with low-momentum interactions?,''
  Nucl.\ Phys.\ A {\bf 763}, 59 (2005)
  [nucl-th/0504043].
  %%CITATION = NUCL-TH/0504043;%%
  %118 citations counted in INSPIRE as of 28 Jul 2013
  
\bibitem{Ramanan:2007bb}
  S.~Ramanan, S.~K.~Bogner, and R.~J.~Furnstahl,
  %``Weinberg Eigenvalues and Pairing with Low-Momentum Potentials,''
  Nucl.\ Phys.\  A {\bf 797}, 81 (2007).

\bibitem{thouless} D.~J.~Thouless, Ann.\ Phys.\ (N.Y.) {\bf 10}, 553
  (1960).

\bibitem{RingSchuck} P.~Ring and P.~Schuck, {\it The Nuclear Many-Body
  Problem} (Springer-Verlag, Berlin, 1980).

\bibitem{fewbody} W.~Gl\"ockle, \textit{The Quantum Mechanical
  Few-Body Problem} (Springer-Verlag, Berlin, 1983).

\bibitem{Fetter_walecka} A.~L.~Fetter and J.~D.~Walecka, {\it Quantum
  Theory of Many-Particle Systems} (McGraw-Hill, New York, 1971).

\bibitem{Bogner:2006vp}
  S.~K.~Bogner, R.~J.~Furnstahl, S.~Ramanan, and A.~Schwenk,
  %``Low-momentum interactions with smooth cutoffs,''
  Nucl.\ Phys.\  {\bf A784} (2007) 79.

%\bibitem{ZuoLombardo2002} W.~Zuo, U.~Lombardo, H.-J.~Schulze,
%  and C.~W.~Shen, Phys.\ Rev.\ C {\bf 66}, 037303 (2002).

\bibitem{HebelerSchwenk2010} 
  % chiral three-nucleon forces and neutron matter
  K.~Hebeler and A.~Schwenk, Phys.\ Rev.\ C {\bf 82}, 014314 (2010).

\bibitem{SadeMelo1993} C.~A.~R.~S\'a~de~Melo, M.~Randeria, and
  J.~R.~Engelbrecht, Phys.\ Rev.\ Lett.\ {\bf 71}, 3202 (1993).

\bibitem{Regal2004} C.~A.~Regal, M.~Greiner, and D.~S.~Jin,
  Phys.\ Rev.\ Lett.\ {\bf 92}, 040403 (2004).
 
\bibitem{Zwierlein2004} M.~W.~Zwierlein, C.~A.~Stan, and
  C.~H.~Schunck, S.~M.~F. Raupach, A.~J.~Kerman, and W.~Ketterle,
  Phys.\ Rev.\ Lett.\ {\bf 92}, 120403 (2004).
 
\bibitem{Shen2002} 
% Screening Effects on $^1S_0$ Pairing in Neutron Matter
  C.~Shen, U.~Lombardo, P.~Schuck, W.~Zuo, and N.~Sandulescu,
  Phys.\ Rev.\ C {\bf 67}, 061302 (2003).

\bibitem{Ku2012} M.~J.~H.~Ku, A.~T.~Sommer, L.~W.~Cheuk, and
  M.~W.~Zwierlein, Science {\bf 335}, 563 (2012).

\bibitem{GorkovMelikBarkhudarov} L.~P.~Gor'kov and
  T.~K.~Melik-Barkhudarov, J.\ Exptl.\ Theoret.\ Phys.\ (U.S.S.R.)
  {\bf 40}, 1452 (1961) [translation: Sov.\ Phys.\ JETP {\bf 13} 1018
    (1961)].

\bibitem{ZimmermannStolz}
% The Mass Action Law in Two-Component Fermi Systems Revisited
  R.~Zimmermann and H.~Stolz,
  Phys. Status Solidi B \textbf{131}, 151 (1985).

\end{thebibliography}
\end{document}